\documentclass[prb,twocolumn]{revtex4-1}
\usepackage{graphicx}
\usepackage{dcolumn}
\usepackage{bm}
\usepackage [latin1]{inputenc}
\usepackage{epstopdf}
\usepackage{color}
\usepackage{ulem}
\usepackage{multirow}
\begin{document}

\title{{Transition to a supersolid phase in a
two-dimensional dilute gas of electron-hole pairs}}

\author{D.\,V.\,Fil$^{1,2}$, S.\,I.\,Shevchenko$^3$}

\email{fil@isc.kharkov.ua}

\affiliation{$^1$Institute for Single Crystals, National Academy of Sciences of Ukraine,
60 Nauky Avenue, Kharkiv 61072, Ukraine\\
$^2$V.N. Karazin Kharkiv National University, 4 Svobody Square, Kharkiv 61022,
Ukraine\\$^3$B.~Verkin Institute for Low Temperature Physics and Engineering,\\ National
Academy of Sciences of Ukraine,\\ Lenin av. 47 Kharkov 61103, Ukraine}

\begin{abstract}

Using coherent-state formalism (the Keldysh formalism), the article describes a
transition from a homogeneous superfluid state to a  supersolid state in a
two-dimensional dilute gas of electron-hole pairs with spatially separated components.
Such a transition is heralded by  the appearance of a roton-type minimum in the
collective excitation spectrum, which touches the abscissa axis as the distance between
the  layers or the pair density increases. This signals the instability of the system
with respect to the appearance of a spatial modulation of  the pair density. It has been
found that a first-order transition to a hexagonal supersolid phase takes place a little
earlier. A theory without  phenomenological constants has been developed for an arbitrary
relation between the effective masses of an electron and a hole. A phase  diagram for the
system has been plotted in the variables "the chemical potential of pairs - the distance
between the layers". It has been shown  that there is a jump in the average density of
the condensate during the phase transition. It has been established that with an increase
in the  chemical potential, the inhomogeneous phase breaks up into high-density regions
surrounded by lines at which the density becomes zero,  with these lines forming a
continuous network.
   \end{abstract}

\maketitle

\section{INTRODUCTION}

A supersolid phase is a state that combines superfluid properties with crystalline order.
This possibility, as applied to quantum  crystals of solid $^4$He, was predicted by A. F.
Andreev and I. M.  Lifshitz \cite{1}. There has been renewed interest in this phenomenon
due  to experiments that have revealed a decrease in the oscillation period of a torsion
pendulum filled with solid helium, and which were repeatedly reproduced in different
laboratories (see, for example, reviews in Refs. \onlinecite{2} and \onlinecite{3}). The
observed effect  could be attributed to the appearance of a superfluid fraction,  which
is not involved in torsional oscillations, but additional experiments in Ref.
\onlinecite{4} and the theory in Refs. \onlinecite{5,6,7} suggest that this effect is
caused not by the appearance of the supersolid phase, but rather  by superplasticity.

A supersolid state can arise not only in quantum crystals. As  was shown in Ref.
\onlinecite{8} (see also Refs. \onlinecite{9, 10,11}), Bose gases with dipole interaction
between particles belong to systems in which superfluid properties and spatial
periodicity can be expected to coexist. The supersolid state in dipole quantum gases was
experimentally discovered recently in studies undertaken by three different groups
\cite{12,13,14}. The interaction between dipole particles comprises a long-range
component. The Fourier component of the interaction  potential $V_\mathbf{k}$ is a
function of the wave vector. In a Bose condensate state, the excitation spectrum
$\omega(k)$ is described by an equation similar to the one for the Bogolyubov spectrum,
with the difference  being that the interaction constant is replaced with $V_\mathbf{k}$:
$\omega(k)=\sqrt{\varepsilon_k(\varepsilon_k+2 V_k n)}$, where $\varepsilon_k$ is the
kinetic energy of the particles and $n$ is the condensate density (it is assumed that
$V_\mathbf{k}$ depends only on the modulus $\mathbf{k}$). If in a certain range of wave
vectors the Fourier component $V_{k}$ takes negative values and the inequality
$\varepsilon_k+2 n V_k<0$ is achieved, then a spatially uniform condensate is unstable.

The authors of Ref. \onlinecite{8} show that various inhomogeneous  phases can arise in
dipole quantum gases, depending on the value of the dimensionless parameter composed of
the constants of two-particle and three-particle contact interactions, condensate
density, and the dipole moment magnitude. Such phases include a one-dimensional
supersolid phase (in the form of stripes), a two-dimensional supersolid phase with a
triangular lattice, as well as a  phase with a honeycomb-type lattice (similar to a
graphene lattice).  In Ref. \onlinecite{8}, the discussion pertains to dipole molecules.
Therefore, the  temperature at which a transition from a homogeneous superfluid state to
a supersolid state can occur is very low (no more than several tens of nK), which is
consistent with the temperature (T=20 nK) at which the supersolid phase was observed
experimentally \cite{13}.

Due to the long-discussed possibility of superfluidity of  coupled electron-hole pairs
with spatially separated components (see Ref. \onlinecite{15}), a transition to the
supersolid phase in bilayer  electron-hole systems has also been considered. A pioneering
study  on electron-hole pairing in quantum Hall systems \cite{16} showed that the
collective excitation spectrum contains a roton-type minimum.  It was found that when a
critical distance between layers is reached,  the minimum point touches the abscissa axis
and a phase transition  should be observed in the system. In Ref. \onlinecite{17}, a
phase diagram of a bilayer electron-hole system in a zero magnetic field was analyzed
qualitatively. Arguments were made that with increasing distance between the layers, a
phase transition to a supersolid phase should first be observed, which is then followed
by a transition to a  Wigner crystal phase. A similar result was obtained in Ref.
\onlinecite{18},  based on the approach developed by the author of that study and
referred to by him as the Ginzburg-Landau quantum theory. In Ref. \onlinecite{19}, it was
established that in a bilayer electron-hole system with a significant imbalance of
electron and hole densities, the exciton gas condenses into a phase resembling the
Fulde-Ferrel-Larkin-Ovchinnikov phase, which can be considered as a variation  of the
supersolid phase. In this case, the formation of such a phase is promoted by a
considerable difference between the electron and hole effective masses.

Our previous study \cite{20} developed an approach for describing the condensate of
electron-hole pairs in a bilayer system, using the coherent-state formalism proposed by
L. V. Keldysh in relation  to a three-dimensional exciton condensate \cite{21}. As part
of the description \cite{20}, an analytical expression was obtained for the collective
excitation spectrum. It turns out that it is important to take into account not only the
dependence on the wave vector of the Fourier  component of the direct Coulomb interaction
between electron-hole pairs, but also a similar dependence for the exchange interaction.
The spectrum obtained in Ref. \onlinecite{20} (much like the collective excitation
spectrum in dipole Bose gases, as well as the spectrum obtained in Ref. \onlinecite{16}),
has a roton-type minimum, and with an increase in the distance between the layers or in
the pair density, the state with a uniform Bose condensate becomes unstable.

In the present study, the approach developed in Ref. \onlinecite{20} is used to describe
the supersolid phase. We have obtained expressions for  the energy of inhomogeneous
phases. The formalism used does not contain phenomenological parameters. The energy
depends on the  ratio of the distance between the layers $d$ to the effective Bohr radius
of the pair $a_0$, on the chemical potential of the pairs, and the ratio of the effective
masses of the electron ($m_e$) and the hole ($m_h$).  The particular inhomogeneous phase
that corresponds to the  minimum energy has been established, and how the spatial
distribution of the condensate in the inhomogeneous phase changes depending on the
indicated parameters.

\section{THE GENERAL EXPRESSION FOR THE EXCITON
CONDENSATE ENERGY. DERIVATION WITHIN THE FRAMEWORK OF THE COHERENT-STATE FORMALISM}

The coherent-state formalism \cite{21} was further developed in Refs. \onlinecite{22} and
\onlinecite{23}, where it was used to describe electron-hole pairing  in two-layer
quantum Hall systems; in Refs. \onlinecite{24, 25, 26}, where it was  employed to analyze
polarization phenomena in a three-dimensional  superfluid gas of electron-hole pairs
(without spatial separation of  electrons and holes); as well as in Ref. \onlinecite{27},
where this formalism was  used to describe the superfluid state of a dilute gas formed by
alkali  metal atoms.

 Let us outline the approach used. An exciton condensate is
described by the many-particle wave function:
\begin{equation}\label{1}
 |\Phi\rangle=e^{\hat{D}}|0\rangle,
\end{equation}
where
\begin{equation}\label{2}
  \hat{D}=\int d {\bf
    r}_1 d{\bf
    r}_2 \Phi({\bf
    r}_1, {\bf
    r}_2)\psi_e^+({\bf
    r}_1) \psi_h^+({\bf
    r}_2)-H.c.,
\end{equation}
$\psi_e^+$ and $\psi_h^+$ are the electron and hole creation operators, $\Phi({\bf
    r}_1, {\bf r}_2)$ is the
pair wave function in the dilute condensate, the  ${\bf r}_i$ vectors are two-dimensional
(i.e. they lie in the planes of the electron and hole layers), and the wave function
$|0\rangle$ corresponds to a vacuum state (a state in which there are no electrons and
holes). Here we consider a one-component condensate of pairs. As was shown in Ref.
\onlinecite{20} (see also Ref. \onlinecite{28}), a two-component pair condensate (pairs
differ in the projection of the spin of an electron or hole) is unstable with respect to
spatial separation  at  $d/a_0>0.2$. At the same time, according to the estimates below,
the supersolid phase can only occur at  $d/a_0>1$.

The function $\Phi({\bf r}_1, {\bf r}_2)$  is determined from the minimum condition for
the energy of the system. The Hamiltonian of the system is taken as
\begin{eqnarray}\label{3}
    H=-\sum_{\alpha=e,h}\int d {\bf r}\frac{\hbar^2}{2
    m_\alpha}\psi^{+}_{\alpha}({\bf r})\nabla^2
    \psi_{\alpha}({\bf r})\cr
    +\frac{1}{2}\sum_{\alpha,\beta=e,h}
    \int d {\bf
    r}d{\bf
    r'}\psi^{+}_{\alpha}({\bf r})
    \psi^+_{\beta}({\bf r'})
    V_{\alpha\beta}(|{\bf r}-{\bf r'}|)
    \psi_{\beta}({\bf r'})\psi_{\alpha}({\bf
    r}),\cr
\end{eqnarray}
where $V_{\alpha\beta}(r)$ is the energy of the Coulomb interaction.  Let us consider a
system in a homogeneous dielectric matrix  with the permittivity $\varepsilon$, which
coincides with the dielectric constant of the insulator that separates the electron and
hole layers. Then  $V_{ee}(r)=V_{hh}(r)=e^2/\varepsilon r$ and
$V_{eh}(r)=-e^2/\varepsilon \sqrt{r^2+d^2}$. The operators of the total number of
electrons and holes are written as follows: $ \hat{N}_e=\int d {\bf r} {\psi}^+_{e}({\bf
r}){\psi}_{e}({\bf r})$, $\hat{N}_h=\int d {\bf r} {\psi}^+_{h}({\bf r}) {\psi}_{h}({\bf
r})$.

For further analysis it is convenient to move on to the operators
$\tilde{\psi}_{\alpha}({\bf r})=e^{-\hat{D}}\psi_{\alpha}({\bf r})e^{\hat{D}}$
($\alpha=e,h$), which are expressed in terms of the creation and annihilation operators
for electrons and holes \cite{21}:
\begin{eqnarray}
\label{4}
  \tilde{\psi}_{e}({\bf r}) &=& \int d {\bf r'}[C_e({\bf r},{\bf r'})
  {\psi}_{e}({\bf r'})
  +S({\bf r},{\bf r'}){\psi}^+_{h}({\bf r'})], \cr
  \tilde{\psi}_{h}({\bf r}) &=&
  \int d {\bf r'}[C_h({\bf r'},{\bf r}){\psi}_{h}({\bf r'})
  -S({\bf r'},{\bf r}){\psi}^+_{e}({\bf
  r'})],
\end{eqnarray}
where
\begin{eqnarray}\label{5}
 C_e({\bf r},{\bf r'}) &=& \delta(\mathbf{r}-\mathbf{r}')
 +\sum_{n=1}^\infty \frac{(-1)^n}{(2n)!}
 (\Phi\cdot\Phi^+)^n,\cr
 C_h({\bf r},{\bf r'})&=& \delta(\mathbf{r}-\mathbf{r}')
 +\sum_{n=1}^\infty \frac{(-1)^n}{(2n)!}
 (\Phi^+\cdot\Phi)^n,\cr
  S({\bf r},{\bf r'}) &=& \sum_{n=0}^\infty\frac{(-1)^n}{(2n+1)!}\Phi \cdot
  (\Phi^+\cdot\Phi)^n.
\end{eqnarray}
The explicit form of the Hermitian-conjugate operators   $\tilde{\psi}_{\alpha}^+({\bf
r})=e^{-\hat{D}}\psi_{\alpha}^+({\bf r})e^{\hat{D}}$ is determined by Hermitian-conjugate
Eq. (\ref{4}).  Equation (\ref{5}) uses the notation $\Phi^+({\bf r}_1,{\bf
r}_2)\equiv\Phi^*({\bf r}_2,{\bf r}_1)$  and the multiplication sign denotes convolution.
Using Eq. (\ref{4}) and (\ref{5}), we can express the energy of the system $
  E=\langle\Phi|H|\Phi\rangle=\langle 0|\tilde{H}|0\rangle$,
  the total number of electrons $
 N_e=\langle\Phi|\hat{N}_e|\Phi\rangle=\langle
 0|\tilde{{N}_e}|0\rangle $ and
the total number of holes $
 N_h=\langle\Phi|\hat{N}_h|\Phi\rangle=\langle
 0|\tilde{{N}_h}|0\rangle $ in terms of the
operators $\tilde{H}$, $\tilde{{N}_e}$ and $\tilde{{N}_h}$. The latter are obtained from
Hamiltonian (\ref{3}) and the operators   $\hat{N}_e$ and $\hat{N}_h$ by replacing
$\psi_{e(h)}$ in them with $\tilde{\psi}_{e(h)}$, and by making a similar replacement for
the Hermitian-conjugate operators. It is easy to show that in the state (\ref{1}), the
total number of electrons is  equal to the total number of holes, $N_e=N_h$.

Let us determine how the state of the system changes depending on the chemical potential
of pairs $\mu=(\mu_e+\mu_h)/2$, which is considered as an external parameter. The desired
state corresponds to $\Phi({\bf r}_1, {\bf r}_2)$, at which the minimum potential
$\Omega=E-\mu N$ is reached, where $N=N_e=N_h$is the number of pairs.

At low pair density (when the average distance between pairs is much larger than the pair
size), the $\Omega$ function can be written as a series in powers of $\Phi$. We shall
confine ourselves to taking into  account the terms of the second and fourth order in
$\Phi$. In this  approximation, we obtain the following expression:
\begin{eqnarray}
\label{6}
  \Omega=\int d\mathbf{r}_1 d\mathbf{r}_2  d\mathbf{r}_3 d\mathbf{r}_4 \Bigg\{
  \Bigg[\delta(\mathbf{r}_2-\mathbf{r}_4)\delta(\mathbf{r}_1-\mathbf{r}_3)
  \cr -\frac{1}{3}\Phi^+({\bf
r}_2, {\bf r}_3)\Phi({\bf r}_3, {\bf r}_4)\Bigg]\cr\times\Phi^+({\bf r}_4, {\bf r}_1)
    \Bigg(-\frac{\hbar^2}{2 m_e}\nabla_{\mathbf{r}_1}^2
    -\frac{\hbar^2}{2 m_h}\nabla_{\mathbf{r}_2}^2-\mu\cr +V_{eh}(|{\bf r}_1- {\bf r}_2|)\Bigg)
\Phi({\bf r}_1, {\bf r}_2)\cr +\frac{1}{2}R_d(\mathbf{r}_1, \mathbf{r}_2, \mathbf{r}_3,
\mathbf{r}_4)\cr \times\Phi^+({\bf r}_2, {\bf r}_1)\Phi({\bf r}_1, {\bf r}_2)\Phi^+({\bf
r}_4, {\bf r}_3)\Phi({\bf r}_3, {\bf r}_4)\cr - \frac{1}{2}R_{ex}(\mathbf{r}_1,
\mathbf{r}_2, \mathbf{r}_3, \mathbf{r}_4)\cr \times \Phi^+({\bf r}_2, {\bf r}_1)\Phi({\bf
r}_1, {\bf r}_4)\Phi^+({\bf r}_4, {\bf r}_3)\Phi({\bf r}_3, {\bf r}_2) \Bigg\},
\end{eqnarray}
where
\begin{eqnarray}
\label{7} R_d(\mathbf{r}_1, \mathbf{r}_2, \mathbf{r}_3, \mathbf{r}_4)=V_{ee}(r_{13})+
V_{hh}(r_{24})\cr +V_{eh}(r_{14})+V_{eh}(r_{23}),\cr R_{ex}(\mathbf{r}_1, \mathbf{r}_2,
\mathbf{r}_3, \mathbf{r}_4)=V_{ee}(r_{13})+ V_{hh}(r_{24})\cr
+\frac{1}{2}\Bigg[V_{eh}(r_{14})+V_{eh}(r_{23})+V_{eh}(r_{12}) +V_{eh}(r_{34})\Bigg].
\end{eqnarray}
Equation (\ref{7}) use the notation $r_{ik}=|\mathbf{r}_i-\mathbf{r}_k|$; indices 1 and 3
relate to electrons, and indices 2 and 4 to holes.

The function  $\Phi({\bf r}_1, {\bf r}_2)$ is determined as follows:
\begin{equation}\label{8}
\Phi({\bf r}_1, {\bf r}_2)=\Psi(\mathbf{R}_{12})\phi_0(\mathbf{r}_{12}),
\end{equation}
where $\mathbf{R}_{12}=(m_e\mathbf{r}_1+m_h\mathbf{r}_2)/(m_e+m_h)$ is the coordinate of
the  centre of mass, $\phi_0(\mathbf{r})$ is the wave function of the lowest energy bound
state of the pair. The function $\phi_0(\mathbf{r})$ is determined from the  Schroedinger
equation
\begin{equation}\label{9}
\left[-\frac{\hbar^2}{2 m}\nabla^2_\mathbf{r}
+V_{eh}(r)\right]\phi_0(\mathbf{r})=E_0\phi_0(\mathbf{r}),
\end{equation}
where $m=m_e m_h/(m_e+m_h)$ is the reduced mass, and $E_0$ is the energy of the ground
state of the pair. The function $\phi_0(\mathbf{r})$ is normalized by the condition $\int
d^2 r |\phi_0(\mathbf{r})|^2=1$.

We count the chemical potential from $E_0$ and introduce $\tilde{\mu}=\mu-E_0$. In the
case of a spatially uniform condensate, the function (\ref{8}) is equal to
$\Phi(\mathbf{r}_1,\mathbf{r_2})=\Phi(\mathbf{r}_{12})=\sqrt{n_0}\phi_0(\mathbf{r})$,
where the $n_0$ value is determined from the minimum condition for $\Omega$. In the low
density limit, $n_0$ coincides with the pair density (see below). The  energy (\ref{6})
per unit area takes the following form:
\begin{equation} \label{10}
    \frac{\Omega_u}{S}=-\tilde{\mu}\left(n_0-\frac{1}{3}n_0^2\int
    \frac{d^2 q}{(2\pi)^2}|\phi_\mathbf{q}|^4\right)+\frac{\gamma_0}{2}n_0^2,
\end{equation}
where $S$ is the area of the system, $\phi_\mathbf{q}$ is the Fourier component of  the
function $\phi_0(\mathbf{r})$ (to simplify, we write the Fourier component without index
0), $\gamma_0$ is the interaction constant. This constant is contributed to by direct and
exchange interactions, $\gamma_0=\gamma_{0}^{(d)}+\gamma_{0}^{(ex)}$, where
\begin{equation}\label{11}
   \gamma_{0}^{(d)}=\frac{4\pi e^2 d}{\varepsilon},
\end{equation}
\begin{eqnarray}\label{12}
 \gamma_{0}^{(ex)}= -\frac{4\pi e^2}{\varepsilon}
\int\frac{d^2 p}{(2\pi)^2}\frac{d^2
 q}{(2\pi)^2}\frac{1}{p}|\phi_{\mathbf{q}}|^2\Bigg[
|\phi_{\mathbf{q} +\mathbf{p}}|^2\cr -\frac{e^{- p
d}}{2}\left(\phi^*_{\mathbf{q}+\mathbf{p}}\phi_{\mathbf{q}}+
\phi^*_{\mathbf{q}}\phi_{\mathbf{q}+\mathbf{p}}\right)\Bigg].
\end{eqnarray}

The minimum (\ref{10}) corresponds to
\begin{equation}\label{13}
    n_0=\frac{\tilde{\mu}}{\gamma_0+\frac{2}{3}\tilde{\mu}\int
    \frac{d^2 q}{(2\pi)^2}|\phi_\mathbf{q}|^4}.
\end{equation}

As can be seen from Eq. (\ref{13}), at small $\tilde{\mu}$, the $n_0$ value depends
almost linearly on $\tilde{\mu}$. The inclusion of the second term in the denominator of
Eq. (\ref{13}) gives a correction to $n_0$ of the order of $\tilde{\mu}^2$. Substituting
(\ref{13}) into (\ref{10}), we obtain the energy of the system, which is proportional (in
the lowest approximation) to $\tilde{\mu}^2$:
\begin{equation} \label{14}
    \frac{\Omega_u}{S}= -\frac{1}{2}\frac{\tilde{\mu}^2}{\gamma_0},
\end{equation}
and the correction $\delta n_0\propto \tilde{\mu}^2$ gives a correction to$\Omega_u$,
which is proportional to $\tilde{\mu}^3$. Thus, the inclusion of the second term in the
denominator of Eq. (\ref{14}) is an excess of accuracy because in (\ref{6}) we neglect
the terms proportional to $\Phi^6$, which would also give a correction of the order of
$\tilde{\mu}^3$. Since we restrict ourselves to an approximation which does not take
account of the contribution to the energy from terms of the order of $n_0^3$, we can,
from the outset, neglect the term $\Phi^+({\bf r}_2, {\bf r}_3)\Phi({\bf r}_3, {\bf
r}_4)/3$ in the first round brackets in (\ref{6}).

The quantity $\gamma_0$ is positive. If $\gamma_0$ were negative, there would be  a
collapse, but we do not consider this case here.

In a spatially homogeneous case, it is possible to find the exact relation between $n_0$
and the pair density $n_p$. In this case, Eq. (\ref{4}) written in momentum
representation are reduced to the usual $u-v$ transformation:
\begin{eqnarray}\label{15}
  \tilde{\psi}_{e}({\bf q})  &=& u_\mathbf{q}  {\psi}_{e}({\bf q})
 +v_\mathbf{q}  {\psi}_{h}^+(-{\bf q}),\cr
 \tilde{\psi}_{h}(-{\bf q})  &=& u_\mathbf{q}  {\psi}_{h}(-{\bf q})
 -v_\mathbf{q}  {\psi}_{e}^+({\bf q}),
\end{eqnarray}
where
\begin{equation}\label{16}
u_\mathbf{q}=\cos|\Phi_{\mathbf{q}}|,\quad
v_\mathbf{q}=\frac{\Phi_{\mathbf{q}}}{|\Phi_{\mathbf{q}}|}\sin|\Phi_{\mathbf{q}}|,
\end{equation}
and $\Phi_\mathbf{q}=\sqrt{n_0} \phi_\mathbf{q}$  is the Fourier component of the
function $\Phi(\mathbf{r})$. The pair density is as follows:
\begin{equation}\label{17}
 n_p=\int\frac{d^2 q}{(2\pi)^2} |v_\mathbf{q}|^2=
 \int\frac{d^2 q}{(2\pi)^2}
 \left[\sin\left(\sqrt{n_0}|\phi_\mathbf{q}|\right)\right]^2.
\end{equation}
As can be seen, in the limit of $\Phi_\mathbf{q}\ll 1$, $n_0$ approaches to the pair
density, but in the general case $n_0 > n_p$. In the approximation being considered, the
difference between $n_0$ and $n_p$ is neglected.

\section{ENERGY OF THE INHOMOGENEOUS PHASE}

To describe the supersolid phase, we shall take $\Psi(\mathbf{R})$ as a  spatially
periodic function. We use the same functions as in Ref. \onlinecite{8}.

A one-dimensional supersolid phase (wave) is set by the function
\begin{equation}\label{18}
\Psi_{w}(\mathbf{R})=\sqrt{n_0}\left[\cos\theta+\sqrt{2}\sin\theta\cos (k X)\right].
\end{equation}
The parameter $\theta$ varies from $-\pi/2$ to $\pi/2$.

A hexagonal supersolid phase corresponds to the function
\begin{equation}\label{19}
\Psi_{h}(\mathbf{R})=\sqrt{n_0}\left[\cos\theta
+\sqrt{\frac{2}{3}}\sin\theta\sum_{i=1}^3\cos (\mathbf{k}_i \mathbf{R})\right],
\end{equation}
where $\mathbf{k}_1=(k,0)$, $\mathbf{k}_2=(-k/2,\sqrt{3}k/2)$,
$\mathbf{k}_3=(-k/2,-\sqrt{3}k/2)$. At positive $\theta$ values ($\theta<\pi/2$), the
main maxima of the $\Psi_{h}^2(\mathbf{R})$ function form a triangular lattice. At small
negative $\theta$ values, a honeycomb-type lattice emerges. In the low density limit,
$n_0$ corresponds to the average pair density with respect to both the phase (\ref{18})
and the phase (\ref{19}).

By substituting (\ref{18}) and (\ref{19}) into (\ref{6}), we find the $n_0$ value
corresponding to the minimum $\Omega$ at given  $\theta$ and $k$:
\begin{equation}\label{20}
n_0=\frac{\tilde{\mu}-\epsilon_k \sin^2\theta}
    {\gamma_{w,h}(k,\theta)}\Theta\left[\frac{\tilde{\mu}-\epsilon_k \sin^2\theta}
    {\gamma_{w,h}(k,\theta)}\right],
\end{equation}
where $\epsilon_k=\hbar^2 k^2/2(m_e+m_h)$ is the kinetic energy of the pair,
$\gamma_{w,h}(\theta,k)$ is the interaction constant that depends on the parameters $k$
and $\theta$ that determine the form of the function $\Psi(\mathbf{R})$, and $\Theta(x)$
is the Heaviside theta function. The appearance of the theta function is associated with
the condition $n_0\geq 0$. As a result, we obtain the following expression for the
energy:
\begin{equation} \label{21}
    \frac{\Omega_{w,h}(k,\theta)}{S}=-\frac{1}{2}\frac{\left(\tilde{\mu}
-\epsilon_k \sin^2\theta\right)^2}
    {\gamma_{w,h}(k,\theta)}\Theta\left[\frac{\tilde{\mu}-\epsilon_k \sin^2\theta}
    {\gamma_{w,h}(k,\theta)}\right].
\end{equation}
The general structure of the functions $\gamma_{w}(k,\theta)$ and $\gamma_{h}(k,\theta)$
is given by the following equations:
\begin{equation}\label{22}
   \gamma_{w}(k,\theta)=\gamma_0+4\gamma_{2}(k) \cos^2\theta \sin^2\theta
   +\gamma_{4,w}(k) \sin^4\theta,
\end{equation}
\begin{eqnarray}\label{23}
   \gamma_{h}(k,\theta)=\gamma_0+4\gamma_{2}(k) \cos^2\theta \sin^2\theta
\cr +\gamma_{3,h}(k)\cos\theta \sin^3\theta +\gamma_{4,h}(k)\sin^4\theta,
\end{eqnarray}
where $\gamma_0$ is the above introduced interaction constant for the homogeneous phase.
Explicit expressions for the functions $\gamma_2(k)$ and $\gamma_{3(4),w(h)}(k)$ are
rather cumbersome, so they are given in the  Appendix.

 At $\theta=0$ Eq. (\ref{21}) turns into
(\ref{14}), i.e. (\ref{21}) also describes the homogeneous phase. The expansion of
(\ref{21}) at small $\theta$ is as follows:
\begin{equation}\label{24}
   \frac{\Omega_{w,h}(k,\theta)}{S}=-\frac{1}{2}\frac{\tilde{\mu}^2}
    {\gamma_0}+\frac{\tilde{\mu}}{\gamma_0}\theta^2
    \left(\epsilon_k+2\frac{\tilde{\mu}}{\gamma_0}\gamma_2(k)\right).
\end{equation}

Given that for the homogeneous phase $n_0=\tilde{\mu}/\gamma_0$, this phase has a higher
energy compared to any of the inhomogeneous phases under consideration, if for some $k$
the inequality $\epsilon_k+2\gamma_2(k)n_0<0$ is satisfied. The latter condition
coincides exactly with the instability condition of the homogeneous state, which follows
from the explicit expression for the collective excitation spectrum \cite{20} (the
condition under which the spectrum becomes imaginary). Even if this condition is not
satisfied, i.e. $\epsilon_k+2 n_0\gamma_2(k)>0$ at all $k$, an inhomogeneous phase at
some finite $\theta$  may have a lower energy compared  to the homogeneous phase. Then a
transition from the homogeneous phase to the supersolid phase will be a first-order phase
transition.

To find an analytical expression for the energy (\ref{21}), we approximate the function
$\phi_0(\mathbf{r})$ by the wave function of the ground state of  a two-dimensional
harmonic oscillator. This approximation is justified for $d> a_0$, where $a_0=\hbar^2
\varepsilon/m e^2$ is the effective Bohr radius of the pair. In this case, the
interaction potential in Eq. (\ref{9}) can be  replaced by its expansion near $r = 0$:
$V_{eh}(r)\approx -e^2/\varepsilon d+e^2r^2/2 \varepsilon d^3$. This results in
$\phi_0({\bf
 r})=({1}/{\sqrt{\pi}r_0})\exp(-{r^2}/{2 r_0^2})$, where $r_0=\sqrt[4]{a_0d^3}$
is the characteristic pair size. The Fourier component of this function is as follows:
$\phi_\mathbf{q}=\sqrt{4\pi}r_0\exp\left(-{q^2 r_0^2}/{2}\right)$. It is convenient to
use the following as a unit of energy:
$$\Xi_0=\frac{e^2}{\varepsilon a_0}=\frac{m
e^4}{\varepsilon^2\hbar^2}$$ (this is the doubled effective Rydberg). Then the energy
$\Omega$ related to  the area $a_0^2$, is given by the expression
\begin{eqnarray}\label{25}
 \tilde{\Omega}_{w,h}(k,\theta) = \frac{\Omega_{w,h} a_0^2}{S}\cr =-\frac{\mathcal{E}_0 }{8 \pi}
 \frac{a_0}{r_0}\frac{\left[\frac{\tilde{\mu}}{\mathcal{E}_0}
 -\frac{k^2 a_0^2}{8}(1-x^2)\sin^2\theta\right]^2}
    {\tilde{\gamma}_{w,h}\left(\tilde{k},\theta\right)}\cr \times
\Theta\left[ \frac{\frac{\tilde{\mu}}{\mathcal{E}_0}-\frac{k^2
a_0^2}{8}(1-x^2)\sin^2\theta}
    {\tilde{\gamma}_{w,h}\left(\tilde{k},\theta\right)}\right],
\end{eqnarray}
where $\tilde{k}=k r_0$ and to describe the electron-hole asymmetry, we introduce the
parameter $x=(m_e-m_h)/(m_e+m_h)$ (the case $x = 0$ corresponds to $m_e=m_h$, and the
limit $x\to -1$ corresponds to infinitely heavy holes). The functions
$\tilde{\gamma}_{w,h}(\tilde{k},\theta)$ have the same structure as expressions
(\ref{22}), (\ref{23}) and depend on $x$ and  $\tilde{d}=d/r_0$, same as on parameters.
The explicit form of these functions is given in the Appendix. It should be noted that
$\tilde{\gamma}_{w,h}(\tilde{k},\theta)$ do not change when $x$  changes its sign.

 By
minimizing the energy (\ref{25}) with respect to $k$ and $\theta$, we find the ground
state of the system. If the minimum is reached at  $\theta=0$ (in this case, (\ref{25})
does not depend on $k$), the ground state corresponds to a homogeneous condensate. This
state is achieved at  low $\tilde{\mu}$ ($\tilde{\mu}>0$). As $\tilde{\mu}$ increases, at
certain critical value of $\tilde{\mu}$  the global minimum jumps to the point with
$\theta_h\ne 0$ and  $k\ne 0$. This minimum corresponds to a hexagonal phase. Our
analysis shows that $\theta_h$  falls in the range $(0,\pi/2)$, and the maxima of the
condensate density form a triangular lattice. The lattice parameter is as follows:
$a_h=4\pi/(\sqrt{3}k)$. In a transition from a homogeneous state to a hexagonal phase,
the average condensate density increases abruptly.

\section{RESULTS AND DISCUSSIONS}

\begin{figure}
\begin{center}
\includegraphics[width=8cm]{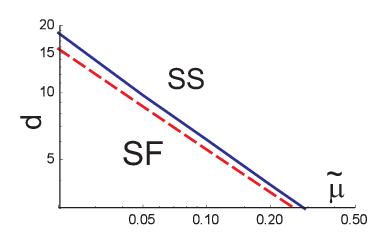}
\end{center}
\caption{Phase diagram in the coordinates "chemical potential - distance between layers".
SF is the homogeneous superfluid phase, SS is the supersolid phase; solid and dashed
lines represent phase transition lines for Systems 1 and 2, respectively, the distance
between the layers $d$ is given in units of $a_0$, the chemical potential $\tilde{\mu}$
in units of $\mathcal{E}_0$.} \label{f1}
\end{figure}

We will not analyze the situation with an arbitrary relation  between $m_e$ and $m_h$,
but will restrict ourselves to two cases. The  first corresponds to the system
 MoS$_2$-MoTe$_2$  (System 1) for  which $m_e=0.47 m_0$ è $m_h=0.62 m_0$, where $m_0$ is the free
electron mass. In this case, $|x|=0.14$. Considering this system in a hexagonal boron
nitride matrix $\varepsilon=5$, we obtain the effective  Bohr radius $a_0\approx 1$ nm.
The second case corresponds to an AlGaAs-based heterostructure (System 2), for which
$m_e=0.067 m_0$, $m_h=0.45 m_0$ and $\varepsilon=13$. For this case, $|x|=0.74$ and
$a_0\approx 12$ nm.  Our analysis shows that over the entire range of parameters  where
the inhomogeneous phase is achieved, the energy of the  phase (\ref{19}) is lower than
the energy of the phase (\ref{18}). In this case, the transition from the homogeneous
phase to phase (\ref{19}) is accompanied by a jump in $\theta$ and a jump in $n_0$, i.e.
it is a first-order phase transition. Further, when referring to the inhomogeneous phase,
we mean phase (\ref{19}). Figure \ref{f1} shows a phase diagram of the system in
variables ($\tilde{\mu}/{\cal E}_0,d/a_0$). It represents phase transition lines for
Systems 1 and 2. With an increase in the parameter $|x|$, the phase transition occurs at
a lower value of $d/a_0$. In the limit of $m_e/m_h\to 0$, the inhomogeneous phase becomes
energetically favorable at all $d/a_0$ (in the low pair density limit).  However, as can
be seen in Fig. 1, a wide variation in the ratio of the electron and hole masses has a
weak effect on the position of the phase transition line.

\begin{figure}
\begin{center}
\includegraphics[width=8cm]{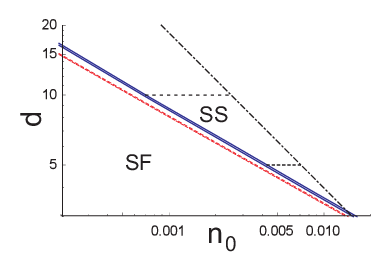}
\end{center}
\caption{ Phase boundary lines in the coordinates "average condensate density - distance
between layers". Solid and dashed lines represent phase transition lines for Systems 1
and 2 respectively. There is a density jump at the phase transition point (see Fig.
\ref{f3}) and therefore, the solid and dashed lines are double, which can be seen at high
resolution. The dash-dotted line limits the range of applicability of the low-density
approximation; $d$ is given in units of $a_0$, and $n_0$ - in units of a $a_0^{-2}$. The
thin dashed lines show the lines along which we calculated the average number of pairs
per cell and the parameter $\theta_h$ as functions of $n_0$ (see Figs. \ref{f4} and
\ref{f5}).} \label{f2}
\end{figure}

Figure \ref{f2} shows phase boundary lines in the coordinates $(n_0 a_0^2,d/a_0)$, and an
imaginary line that outlines the region of the phase diagram where the low-density
approximation is applicable. This line is defined by the equation $4\pi n_0 r_0^2=1$,
which limits the region where $\Phi_\mathbf{q}<1$ for all $\mathbf{q}$ [see Eq.
(\ref{17})]. The  lines dividing the region of parameters with a homogeneous and
inhomogeneous condensate are doubled because there is a density jump associated with a
jump in $n_0$ at the phase transition point.

\begin{figure}
\begin{center}
\includegraphics[width=8cm]{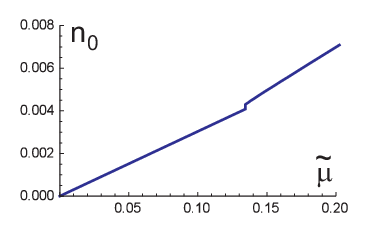}
\end{center}
\caption{ Dependence of $n_0$ (in units of $a_0^{-2}$) on the chemical potential (in
units of $\mathcal{E}_0$) at $d=5 a_0$ for System 1.}\label{f3}
\end{figure}

Figure \ref{f3} shows the dependence of $n_0$ on the chemical potential at $d=5 a_0$  and
$|x|=0.14$. This dependence demonstrates that at the phase transition point, $n_0$
changes abruptly.

\begin{figure}
\begin{center}
\includegraphics[width=8cm]{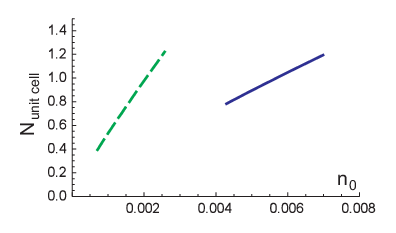}
\end{center}
\caption{
 Average number of pairs per unit cell versus $n_0$ (in units of $a_0^{-2}$) for $d/a_0=5$
 (solid line) and $d/a_0=10$ (dashed line) in System 1.} \label{f4}
\end{figure}

The lattice parameter $a_h$ expressed in terms of $a_0$ increases  with an increase in
the ratio $d/a_0$ approximately according to the law $d^{3/4}$. The same parameter
expressed in terms of $r_0$ varies slightly and remains within $3.5 r_0 < a_h < 5 r_0$
over the entire range of $d$ and $\tilde{\mu}$ values being considered. At a given $d$,
with an increase in the chemical potential and with a corresponding change in the
condensate density, the average number of pairs per unit cell also changes. The
dependence of the average number of pairs per unit cell on $n_0$ is shown in Fig.
\ref{f4}. It can be seen that in the case of the parameters considered here, this number
is less than, or of the order of, one.

\begin{figure}
\begin{center}
\includegraphics[width=8cm]{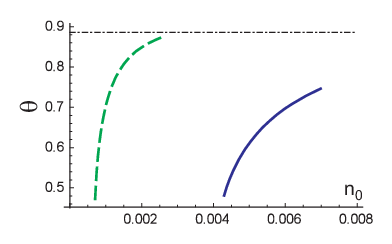}
\end{center}
\caption{ The parameter $\theta_h$ versus $n_0$ (in units of $a_0^{-2}$) for the same $d$
as in Fig. 4. The dash-dotted line represents the $\theta_h$ at which there appears a
network of lines where the local condensate density is zero. } \label{f5}
\end{figure}

\begin{figure}
\begin{center}
\includegraphics[width=6cm]{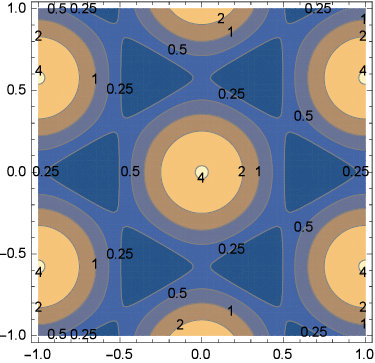}
\end{center}
\caption{Condensate density distribution at $\theta_h=0.48$. Light (yellow-brown) shading
indicates areas with high density; dark (grey-blue) shading indicates areas with low
density. The numbers on the contours represent the values of local density in $n_0$. The
size of the region shown is $4\pi/k\times 4\pi/k$.} \label{f6}
\end{figure}

Figure \ref{f5} shows the dependence of  $\theta_h$ on $n_0$. At a given $d$, the
$\theta_h$ value increases with an increase in $n_0$. The spatial distribution of density
(calculated by the formula $n(\mathbf{r})=|\Psi_{h}(\mathbf{R})|^2$) at $\theta_h\approx
0.48$ is shown in Fig. \ref{f6}. At this  $\theta_h$, a continuous network of regions
with reduced condensate density is formed in the system, thus surrounding density maxima.
In this case, the density of the condensate in the network remains quite high. In an
inhomogeneous condensate, superfluid stiffness will be lower than in a homogeneous
condensate with the same average density (see, for example, Ref. \onlinecite{29}).
Therefore, the temperature of transition from a supersolid phase to  a normal state will
be lower than the temperature of transition from a homogeneous superfluid phase to a
normal state. With increasing $\theta_h$, the density of the condensate in the network
decreases, and at $\theta_h=\arctan(\sqrt{3/2})\approx 0.886$ there appears a continuous
network of lines where the superfluid density becomes zero.

\begin{figure}
\begin{center}
\includegraphics[width=6cm]{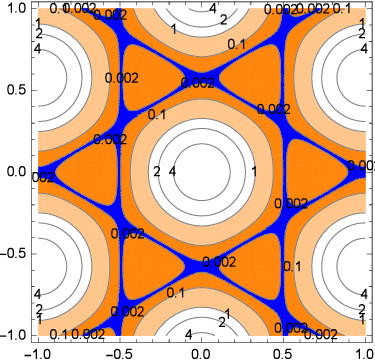}
\end{center}
\caption{The same as in Fig. \ref{f6} for $\theta_h=0.87$. } \label{f7}
\end{figure}

Figure \ref{f7} shows the density distribution at $\theta_h= 0.87$, which is achieved at
$d=10 a_0$ and $4\pi n_0 r_0^2=1$ (see Fig. \ref{f5}). At $\theta_h= 0.87$, a
zero-density network is already nearly formed. The formation of  the network means that
the condensate is divided into a system of  weakly connected regions. Since the average
number of pairs in each region will not be an integer (see Fig. 4), such a state should
collapse already at a low temperature. It can also be assumed that the lattice parameter
ah will adjust to the average density of the condensate so that the number of pairs in
each of the weakly connected regions will become an integer. We leave this for further
study.

Thus, within the coherent-state formalism, we have described the transition of a dilute
gas of electron-hole pairs in a bilayer system to a supersolid state, have plotted a
phase diagram for the system, and have demonstrated how the spatial distribution of the
condensate in this phase changes with a change in the chemical potential.

\section*{APPENDIX: DERIVATION OF EXPLICIT EXPRESSIONS 389
FOR INTERACTION CONSTANTS}

\setcounter{equation}{0}
\renewcommand\theequation{A.\arabic{equation}}

The interaction constants $\gamma_2(k)$ and $\gamma_{3(4),w(h)}(k)$  can be written as
the sum of the terms determined by direct and exchange interactions:
\begin{eqnarray} \label{a0}
  \gamma_2(k) &=& \gamma_2^{(d)}(k) + \gamma_2^{(ex)}(k),\cr
   \gamma_{3(4),w(h)}(k)&=& \gamma_{3(4),w(h)}^{(d)}(k)+\gamma_{3(4),w(h)}^{(ex)}(k)
\end{eqnarray}
($\gamma_{3,w}(k)\equiv 0$).

These terms are expressed in terms of the Fourier transform of the wave function of the
bound state of the pair $\varphi_0(\mathbf{r})$. In $\gamma_2(k)$, the term determined by
direct interaction is as follows:
\begin{eqnarray}\label{a1}
\gamma_2^{(d)}(k)=V_{ee}(k)\int\frac{d^2 p}{(2\pi)^2}\frac{d^2 p'}{(2\pi)^2}
\Big[\phi_{\mathbf{p}+\tilde{m}_h\mathbf{k}} \phi_{\mathbf{p}'-\tilde{m}_h\mathbf{k}}\cr
+ \phi_{\mathbf{p}-\tilde{m}_e\mathbf{k}} \phi_{\mathbf{p}'+\tilde{m}_e\mathbf{k}}
\Big]\phi_\mathbf{p}\phi_{\mathbf{p}'} \cr+V_{eh}(k)\int\frac{d^2 p}{(2\pi)^2}\frac{d^2
p'}{(2\pi)^2} \Big[\phi_{\mathbf{p}+\tilde{m}_h\mathbf{k}}
\phi_{\mathbf{p}'+\tilde{m}_e\mathbf{k}}\cr + \phi_{\mathbf{p}-\tilde{m}_e\mathbf{k}}
\phi_{\mathbf{p}'-\tilde{m}_h\mathbf{k}}\Big]\phi_\mathbf{p}\phi_{\mathbf{p}'},\cr
\end{eqnarray}
where $V_{ee}(k)=2\pi e^2/(\varepsilon k)$, $V_{eh}(k)=-V_{ee}(k)e^{-k d}$ are the
Fourier components of the Coulomb interaction, and the following notation is introduced:
$\tilde{m}_{e(h)}={m}_{e(h)}/(m_e+m_h)$. In what follows, the function $\phi_\mathbf{p}$
is considered real. The contributions of the direct interaction to
$\gamma_{3(4),w(h)}(k)$ can be expressed in terms of the function $\gamma_2^{(d)}(k)$:
\begin{eqnarray} \label{a4}
    \gamma_{3,h}^{(d)}(k)  = 4\sqrt{\frac{2}{3}}\gamma_2^{(d)}(k), \cr
\gamma_{4,w}^{(d)}(k) = \frac{1}{2} \gamma_2^{(d)}(2k),\cr
 \gamma_{4,h}^{(d)}(k) = \frac{2}{3}\left[\gamma_2^{(d)}(k)+\gamma_2^{(d)}
(\sqrt{3}k)\right]+\frac{1}{6}\gamma_2^{(d)}(2k).\cr
\end{eqnarray}
To take into account the contribution of the exchange interaction, we introduce the
function
\begin{eqnarray}
\label{a5} \Gamma(\mathbf{G},\mathbf{g})= -\int\frac{d^2 p}{(2\pi)^2}\frac{d^2
q}{(2\pi)^2}\Bigg\{V_{ee}(q)\cr \times
\Big[\phi_{\mathbf{p}+\mathbf{g}+\tilde{m}_e\mathbf{G}}
\phi_{\mathbf{p}+\mathbf{q}-\mathbf{g}+\tilde{m}_h\mathbf{G}}\cr
+\phi_{\mathbf{p}+\mathbf{q}+\mathbf{g}+\tilde{m}_e\mathbf{G}}
\phi_{\mathbf{p}-\mathbf{g}+\tilde{m}_h\mathbf{G}}\Big]
\phi_{\mathbf{p}+\mathbf{q}+\mathbf{G}} \phi_{\mathbf{p}}\cr +\frac{V_{eh}(q)}{2} \times
\Bigg(\Big[\phi_{\mathbf{p}+\mathbf{q}+\mathbf{g}+\tilde{m}_e\mathbf{G}}
\phi_{\mathbf{p}-\mathbf{g}+\tilde{m}_h\mathbf{G}}\cr
+\phi_{\mathbf{p}+\mathbf{g}+\tilde{m}_e\mathbf{G}}
\phi_{\mathbf{p}+\mathbf{q}-\mathbf{g}+\tilde{m}_h\mathbf{G}}\Big]
\phi_{\mathbf{p}+\mathbf{G}} \phi_{\mathbf{p}}\cr
+\Big[\phi_{\mathbf{p}+\mathbf{q}+\mathbf{g}+\tilde{m}_e\mathbf{G}}
\phi_{\mathbf{p}+\mathbf{q}-\mathbf{g}+\tilde{m}_h\mathbf{G}}\cr
+\phi_{\mathbf{p}+\mathbf{g}+\tilde{m}_e\mathbf{G}}
\phi_{\mathbf{p}-\mathbf{g}+\tilde{m}_h\mathbf{G}}\Big]
\phi_{\mathbf{p}+\mathbf{q}+\mathbf{G}} \phi_{\mathbf{p}}\Bigg)\Bigg\}.
\end{eqnarray}
Expressions for the exchange interaction constants, as written in terms of this function,
take the following form: $$\gamma_0^{(ex)}=\Gamma(0,0)$$,
\begin{eqnarray}\label{a6}
\gamma^{(ex)}_2(k)=\frac{1}{4}\Big[\Gamma(\mathbf{k}_1,0)
+2\Gamma(\tilde{m}_h\mathbf{k}_1,-\tilde{m}_e\tilde{m}_h\mathbf{k}_1)\cr
+2\Gamma(\tilde{m}_e\mathbf{k}_1,\tilde{m}_e\tilde{m}_h\mathbf{k}_1)\cr+
\Gamma((\tilde{m}_e-\tilde{m}_h)\mathbf{k}_1,2\tilde{m}_e\tilde{m}_h\mathbf{k}_1)-
2\Gamma(0,0)\Big],
\end{eqnarray}
\begin{eqnarray}\label{a7}
\gamma^{(ex)}_{3,h}(k)=\sqrt{\frac{2}{3}}\Big\{
\Gamma(\mathbf{k}_1+\tilde{m}_e\mathbf{k}_2,\tilde{m}_e\tilde{m}_h\mathbf{k}_2)\cr
+\Gamma(\mathbf{k}_1 +\tilde{m}_h\mathbf{k}_2,-\tilde{m}_e\tilde{m}_h\mathbf{k}_2) \cr
+\Gamma[\tilde{m}_h\mathbf{k}_1+m_e
\mathbf{k}_2,\tilde{m}_e\tilde{m}_h(\mathbf{k}_1-\mathbf{k}_2)] \cr +
\Gamma[\tilde{m}_e\mathbf{k}_1+\tilde{m}_h\mathbf{k}_2,-\tilde{m}_e\tilde{m}_h
(\mathbf{k}_1-\mathbf{k}_2)]\Big\},
\end{eqnarray}
\begin{eqnarray}\label{a8}
\gamma^{(ex)}_{4,w}(k)=\frac{1}{2}\Big[ \Gamma(2\tilde{m}_h
\mathbf{k}_1,-2\tilde{m}_e\tilde{m}_h\mathbf{k}_1) \cr +\Gamma(2 \tilde{m}_e\mathbf{k}_1
,2\tilde{m}_e\tilde{m}_h\mathbf{k}_1) -\Gamma(0,0)\Big],
\end{eqnarray}
\begin{eqnarray}\label{a9}
\gamma^{(ex)}_{4,h}(k)=\frac{1}{6}\Big[ \Gamma(2\tilde{m}_h
\mathbf{k}_1,-2\tilde{m}_e\tilde{m}_h\mathbf{k}_1) \cr +\Gamma(2 \tilde{m}_e\mathbf{k}_1,
2\tilde{m}_e\tilde{m}_h\mathbf{k}_1) \cr+2\Gamma[\tilde{m}_e(\mathbf{k}_1-\mathbf{k}_2),
\tilde{m}_e\tilde{m}_h(\mathbf{k}_1-\mathbf{k}_2)] \cr +
2\Gamma[\tilde{m}_h(\mathbf{k}_1-\mathbf{k}_2),
-\tilde{m}_e\tilde{m}_h(\mathbf{k}_1-\mathbf{k}_2)] \cr+2\Gamma(\tilde{m}_e\mathbf{k}_1,
,\tilde{m}_e\tilde{m}_h\mathbf{k}_1) \cr +2\Gamma(\tilde{m}_h\mathbf{k}_1
,-\tilde{m}_e\tilde{m}_h\mathbf{k}_1)\cr
+2\Gamma(\mathbf{k}_1+(\tilde{m}_h-\tilde{m}_e)\mathbf{k}_2
,-2\tilde{m}_e\tilde{m}_h\mathbf{k}_2)+2\Gamma(\mathbf{k}_1 \cr +
(\tilde{m}_e-\tilde{m}_h)\mathbf{k}_2 ,2\tilde{m}_e\tilde{m}_h\mathbf{k}_2)
 -5\Gamma(0,0)\Big].
\end{eqnarray}

By substituting the function $\phi_\mathbf{q}=\sqrt{4\pi}r_0\exp(-{q^2 r_0^2}/{2})$ into
the general expressions above we obtain the following result:
\begin{equation}\label{a10}
    \gamma_{w,h}(k,\theta)=\frac{4\pi e^2 r_0}{\varepsilon}
    \tilde{\gamma}_{w,h}(\tilde{k}, \theta),
\end{equation}
where $\tilde{k}=k r_0$  and
\begin{eqnarray}\label{a11}
   \tilde{\gamma}_{w,h}(\tilde{k}, \theta)=\tilde{\gamma}_0
   +4\tilde{\gamma}_{2}(\tilde{k}) \cos^2\theta \sin^2\theta
\cr +\tilde{\gamma}_{3,w(h)}(\tilde{k})\cos\theta \sin^3\theta
+\tilde{\gamma}_{4,w(h)}(\tilde{k})\sin^4\theta
\end{eqnarray}
($\tilde{\gamma}_{3,w}(\tilde{k})\equiv 0$). The $\tilde{\gamma}(\tilde{k})$ functions
included in (\ref{a11}) can be represented as the sum of the contributions of the direct
and exchange interactions, similarly to (\ref{a0}). Explicit expressions for the direct
interaction contributions are equal to $\tilde{\gamma}^{(d)}_0=\tilde{d}$,
\begin{eqnarray}\label{a12}
  \tilde{\gamma}^{(d)}_2(\tilde{k})=\frac{1}{2\tilde{k}}
 \Bigg[\exp\left(-\frac{\tilde{k}^2(1-x)^2}{8}\right)\cr +\exp\left(
 -\frac{\tilde{k}^2(1+x)^2}{8}\right)
 \cr -2\exp\left(-\tilde{k}\tilde{d}-\frac{\tilde{k}^2(1+x^2)}{8}\right)\Bigg],
 \end{eqnarray}
where $\tilde{d}=d/r_0$. We obtain expressions for
$\tilde{\gamma}_{3(4),w(h)}^{(d)}(\tilde{k})$ by replacing,  in (\ref{a4}), all $\gamma$
with $\tilde{\gamma}$, and $k$ with $\tilde{k}$.  To keep a simple record of the exchange
interaction contributions, we determine the functions as follows:
\begin{eqnarray} \label{a13}
  \mathrm{A}(y) =\exp(-y)I_0(y), \cr
  \mathrm{fd}(\tilde{k},\tilde{d})= \sqrt{\frac{2}{\pi}}\int_0^\infty
  \exp\left(-\frac{3p^2}{8}-p \tilde{d}\right)I_0(p\tilde{k})d p,\cr
\end{eqnarray}
where $I_0(y)$ is the modified Bessel function. At $\tilde{k}=0$, the function
$\mathrm{fd}(\tilde{k},\tilde{d})$ can be expressed in terms of the complementary error
function:
\begin{equation}\label{a14}
  \mathrm{fd}(0,\tilde{d})=\mathrm{f_0}(\tilde{d})
  =\sqrt{\frac{4}{3}}
  \exp\left(\frac{2 \tilde{d}^2}{3}\right)
  \mathrm{erfc}\left(\sqrt{\frac{2}{3}}\tilde{d}\right).
\end{equation}
The $\tilde{\gamma}^{(ex)}$ functions written in terms of the functions (\ref{a1}) and
(\ref{a14}) are as follows:
\begin{equation}\label{a15}
\tilde{\gamma}^{(ex)}_0=-\sqrt{\frac{\pi}{2}}\left[1-\mathrm{f_0}(\tilde{d})\right],
\end{equation}
\begin{eqnarray}\label{a16}
\tilde{\gamma}^{(ex)}_2(\tilde{k})=-\frac{1}{4}\sqrt{\frac{\pi}{2}}
\Bigg\{\mathrm{A}\left[\frac{\tilde{k}^2(x-1)^2}{16}\right]\cr \times\left(
\exp\left[-\frac{\tilde{k}^2(x+1)^2}{8}\right]+1\right)\cr+
\mathrm{A}\left[\frac{\tilde{k}^2(x+1)^2}{16}\right]\left(
\exp\left[-\frac{\tilde{k}^2(x-1)^2}{8}\right]+1\right)\cr
+\exp\left[-\frac{\tilde{k}^2(x+1)^2}{8}\right]\left(1-2
\mathrm{fd}\left[\frac{\tilde{k}(x+1)}{4},\tilde{d}\right]\right)\cr+
\exp\left[-\frac{\tilde{k}^2(x-1)^2}{8}\right]\left(1-2
\mathrm{fd}\left[\frac{\tilde{k}|x-1|}{4},\tilde{d}\right]\right)\cr-
\exp\left[-\frac{\tilde{k}^2(x^2+1)}{4}\right]
\left(\mathrm{fd}\left[\frac{\tilde{k}|x|}{2},\tilde{d}\right]
+\mathrm{fd}\left[\frac{\tilde{k}}{2},\tilde{d}\right]\right)\cr-2+2\mathrm{f_0}(\tilde{d})
\Bigg\},\cr
\end{eqnarray}
\begin{eqnarray}\label{a17}
\tilde{\gamma}^{(ex)}_{3,h}(\tilde{k})
=-\sqrt{\frac{4\pi}{3}}\Bigg\{\exp\left[-\frac{\tilde{k}^2(x+1)^2}{8}\right]\cr
\times\mathrm{A} \left[\frac{\tilde{k}^2(x-1)^2}{16}\right]\cr
+\exp\left[-\frac{\tilde{k}^2(x-1)^2}{8}\right] \mathrm{A}
\left[\frac{\tilde{k}^2(x+1)^2}{16}\right]\cr-\exp\left[-\frac{\tilde{k}^2(x^2+1)}{4}\right]
\Bigg(\mathrm{fd}\left[\frac{\tilde{k}\sqrt{1+3x^2}}{4},\tilde{d}\right] \cr
+\mathrm{fd}\left[\frac{\tilde{k}\sqrt{3+x^2}}{4},\tilde{d}\right]\Bigg)\Bigg\},
\end{eqnarray}
\begin{eqnarray}\label{a18}
\tilde{\gamma}^{(ex)}_{4,w}(\tilde{k}) =-\frac{1}{4}\sqrt{\frac{\pi}{2}}\Bigg\{\mathrm{A}
\left[\frac{\tilde{k}^2(x+1)^2}{4}\right]\cr +\mathrm{A}
\left[\frac{\tilde{k}^2(x-1)^2}{4}\right]\cr
+\exp\left[-\frac{\tilde{k}^2(x+1)^2}{2}\right]\cr \times\left(1-2
\mathrm{fd}\left[\frac{\tilde{k}(x+1)}{2},\tilde{d}\right]\right)\cr +
\exp\left[-\frac{\tilde{k}^2(x-1)^2}{2}\right]\cr\times\left(1-2
\mathrm{fd}\left[\frac{\tilde{k}|x-1|}{2},\tilde{d}\right]\right)
-2\left[1-\mathrm{f_0}(\tilde{d})\right]\Bigg\},
\end{eqnarray}
\begin{eqnarray}\label{a19}
\tilde{\gamma}^{(ex)}_{4,h}(\tilde{k})
=\frac{1}{3}\tilde{\gamma}^{(ex)}_{4,w}(\tilde{k})-{\frac{2}{3}}\tilde{\gamma}^{(ex)}_0
\cr -\frac{1}{6}\sqrt{\frac{\pi}{2}}\Bigg\{\mathrm{A}
\left[\frac{3\tilde{k}^2(x+1)^2}{16}\right]
\left(1+\exp\left[-\frac{\tilde{k}^2(x-1)^2}{8}\right]\right)\cr
+\mathrm{A}\left[\frac{3\tilde{k}^2(x-1)^2}{16}\right]
\left(1+\exp\left[-\frac{\tilde{k}^2(x+1)^2}{8}\right]\right)\cr
+\mathrm{A}\left[\frac{\tilde{k}^2(x+1)^2}{16}\right]
\left(1+\exp\left[-\frac{3\tilde{k}^2(x-1)^2}{8}\right]\right)\cr
+\mathrm{A}\left[\frac{\tilde{k}^2(x-1)^2}{16}\right]
\left(1+\exp\left[-\frac{3\tilde{k}^2(x+1)^2}{8}\right]\right)\cr
+\exp\left[-\frac{3\tilde{k}^2(x+1)^2}{8}\right]
\left(1-2\mathrm{fd}\left[\frac{\sqrt{3}\tilde{k}(x+1)}{4},\tilde{d}\right]\right)\cr
+\exp\left[-\frac{3\tilde{k}^2(x-1)^2}{8}\right]
\left(1-2\mathrm{fd}\left[\frac{\sqrt{3}\tilde{k}|x-1|}{4},\tilde{d}\right]\right)\cr
+\exp\left[-\frac{\tilde{k}^2(x+1)^2}{8}\right]
\left(1-2\mathrm{fd}\left[\frac{\tilde{k}(x+1)}{4},\tilde{d}\right]\right)\cr
+\exp\left[-\frac{\tilde{k}^2(x-1)^2}{8}\right]
\left(1-2\mathrm{fd}\left[\frac{\tilde{k}|x-1|}{4},\tilde{d}\right]\right)\cr -
2\exp\left[-\frac{\tilde{k}^2(1+x+x^2)}{2}\right]
\mathrm{fd}\left[\frac{\tilde{k}\sqrt{1+x+x^2}}{2},\tilde{d}\right]\cr
-2\exp\left[-\frac{\tilde{k}^2(1-x+x^2)}{2}\right]
\mathrm{fd}\left[\frac{\tilde{k}\sqrt{1-x+x^2}}{2},\tilde{d}\right]\Bigg\}.\cr
\end{eqnarray}


\begin{thebibliography}{1}
\bibitem{1} A. F. Andreev and I. M. Lifshitz, Zh. Eksp. Teor. Fiz. 56, 2057 (1969) [JETP 29,
 1107 (1969).

\bibitem{2} S. Balibar, Nature 464, 176 (2010).

\bibitem{3} A. B. Kuklov, N. V. Prokof'ev, and B. V. Svistunov, Physics 4, 109 (2011).

\bibitem{4} D. Y. Kim and M. H. W. Chan, Phys. Rev. Lett. 109, 155301 (2012)

\bibitem{5}  A. B. Kuklov, L. Pollet, N. V. Prokof'ev, and B. V. Svistunov, Phys. Rev. B 90,
444 184508 (2014).

\bibitem{6} A. B. Kuklov, Phys. Rev. B 100, 014513 (2019).

\bibitem{7} A. B. Kuklov, N. V. Prokof'ev, and B. V. Svistunov, Fiz. Nizk. Temp. 46, 549
(2020) [Low Temp. Phys. 46, No. 5 (2020)].

\bibitem{8}
ZZ.-K. Lu, Y. Li, D. S. Petrov, and G. V. Shlyapnikov, Phys. Rev. Lett. 115, 075303
(2015).

\bibitem{9} F. Cinti and M. Boninsegni,  Phys. Rev. A 96, 013627 (2017).


\bibitem{10} S. V. Andreev,
Phys. Rev. B 95, 184519 (2017).

\bibitem{11} D. Baillie and P.B. Blakie,  Phys. Rev. Lett. 121, 195301  (2018).

\bibitem{12} L. Tanzi, E. Lucioni, F. Fama, J. Catani, A. Fioretti, C. Gabbanini,
R.N. Bisset, L. Santos, and G. Modugno,  Phys. Rev. Lett. 122, 130405 (2019).


\bibitem{13} F. Bottcher, J.-N. Schmidt,
M. Wenzel, J. Hertkorn, M. Guo, T. Langen, and T. Pfau,  Phys. Rev. X 9, 011051 (2019).


\bibitem{14} L. Chomaz, D. Petter, P. Ilzhofer, G. Natale, A. Trautmann, C. Politi,
G. Durastante, R.M.W. van Bijnen, A. Patscheider, M. Sohmen, M.J. Mark, and F. Ferlaino,
 Phys. Rev. X 9,
021012 (2019).

\bibitem{15} D. V. Fil and S. I. Shevchenko, Fiz. Nizk. Temp. 44, 1111 (2018) [Low Temp.
Phys. 44, 867 (2018)].

\bibitem{16}  H. A. Fertig,  Phys. Rev. B 40, 1087 (1989).

\bibitem{17} Y. N. Joglekar, A. V. Balatsky, and S. Das Sarma,
 Phys. Rev. B 74, 233302 (2006).

\bibitem{18} J.Ye,  Journal of Low Temperature Physics 158,
882 (2010).

\bibitem{19} M. M. Parish, F. M. Marchetti and P. B. Littlewood,  EPL 95, 27007 (2011).


\bibitem{20} D. V. Fil and S. I. Shevchenko, Fiz. Nizk. Temp. 42, 1013 (2016) [Low Temp.
Phys. 42, 794 (2016)].

\bibitem{21} L. V. Keldysh, "Coherent states of exitons", in Problems of Theoretical Physics.
 In Memory of Igor Evgenievich Tamm, edited by V. I. Ritus (Nauka, Moscow, 1972), p.
433; Phys. Usp. 87, 1273 (2017).

\bibitem{22} A. I. Bezuglyj, S. I. Shevchenko, Phys. Rev. B
75, 075322 (2007).

\bibitem{23} A. I. Bezuglyj and S. I. Shevchenko, Fiz. Nizk. Temp. 35, 479 (2009) [Low
Temp. Phys. 35, 373 (2009)]

\bibitem{24} S. I. Shevchenko and A. S. Rukin, Pis'ma v ZhETF 90, 46 (2009) [JETP Lett.
 90, 42 (2009)].

\bibitem{25}S. I. Shevchenko and A. S. Rukin, Fiz. Nizk. Temp. 36, 186 (2010) [Low Temp.
 Phys. 36, 146 (2010)].

\bibitem{26} S. I. Shevchenko and A. S. Rukin, Fiz. Nizk. Temp. 36, 748 (2010) [Low Temp.
Phys. 36, 596 (2010)].

\bibitem{27} S. I. Shevchenko and A. S. Rukin, Fiz. Nizk. Temp. 38, 1147 (2012) [Low
Temp. Phys. 38, 905 (2012)].

\bibitem{28} F.-C. Wu, F. Xue, and A.H. MacDonald,  Phys. Rev. B 92, 165121 (2015).

\bibitem{29} N. Sepulveda, C. Josserand  S. Rica,
 Eur. Phys. J. B 78, 439
(2010).
\end{thebibliography}
\end{document}